\documentclass[12pt]{article}

\usepackage[left=2.8cm,right=2.8cm,top=2.8cm,bottom=2.8cm]{geometry}

\usepackage[colorlinks=true,linkcolor=red,citecolor=blue,urlcolor=green,filecolor=black]{hyperref}

\usepackage{tensor}
\usepackage{amstext,amssymb,amsfonts,amsmath,amsthm}
\usepackage{textcomp}
\usepackage{mathrsfs}
\usepackage{multirow}
\usepackage{array}
\usepackage{color}
\usepackage[nosort]{cite}

 \csname
@addtoreset\endcsname{equation}{section}

\def\be{\begin{equation}}
\def\ee{\end{equation}}
\def\bea{\begin{eqnarray}}
\def\eea{\end{eqnarray}}
\def\ba{\begin{array}}
\def\ea{\end{array}}
\def\bwt{\begin{widetext}}
\def\ewt{\end{widetext}}

\newcommand{\f}[2]{\frac{#1}{#2}}
\newcolumntype{C}[1]{>{\centering}m{#1}}

\begin{document}


\vspace{30pt}

\begin{center}


{\Large\sc Timelike duality, $M'$-theory and an exotic form of the Englert solution} 


\vspace{25pt}
{\sc Marc Henneaux${}^{\, a}$ and Arash Ranjbar${}^{\, a, b}$}

\vspace{10pt}
{${}^a$\sl\small
Universit\'e Libre de Bruxelles and International Solvay Institutes, ULB-Campus Plaine CP231, B-1050 Brussels, Belgium\\
\vspace{10pt}
${}^b$\sl \small Centro de Estudios Cient\'{i}ficos (CECs), Av. Arturo Prat 514, Valdivia, Chile}\\
\vspace{15pt}
\sl \small Email: henneaux@ulb.ac.be, aranjbar@ulb.ac.be

\vspace{40pt}

{\sc\large Abstract} 
\end{center}
\noindent

Through timelike dualities, one can generate exotic versions of $M$-theory with different spacetime signatures.  These are the  $M^*$-theory with signature $(9,2,-)$, the $M'$-theory, with signature $(6,5,+)$ and the theories with reversed signatures $(1,10, -)$, $(2,9, +)$ and $(5,6, -)$.  In $(s,t, \pm)$, $s$ is the number of space directions, $t$ the number of time directions, and  $\pm$ refers to the sign of the kinetic term of the $3$ form.  

The only irreducible pseudo-riemannian manifolds admitting absolute parallelism are, besides Lie groups,  the seven-sphere $S^7 \equiv SO(8)/SO(7)$  and its pseudo-riemannian version $S^{3,4} \equiv SO(4,4)/SO(3,4)$.  [There is also the complexification $SO(8,\mathbb{C})/SO(7, \mathbb{C})$, but it is of dimension too high for our considerations.] The seven-sphere $S^7\equiv S^{7,0}$ has been found to play an important role in $11$-dimensional supergravity, both through the Freund-Rubin solution and the Englert solution that uses its remarkable parallelizability to turn on non trivial internal fluxes.  The spacetime manifold is in both cases $AdS_4 \times S^7$.  We show that $S^{3,4}$ enjoys a similar role in $M'$-theory and construct the exotic form $AdS_4 \times S^{3,4}$ of the Englert solution, with non zero internal fluxes turned on.  There is no analogous solution in $M^*$-theory. 
  

\newpage


\tableofcontents



\newpage

\section{Introduction}

Duality suggests that even the spacetime signature is a relative concept that depends on the description \cite{Hull:1998ym, Hull:1998vg}.  The change of signature induced by duality transformations can be given a group theoretical interpretation in terms of the infinite-dimensional  Kac-Moody algebras $E_{11}$ or $E_{10}$ that have been conjectured to be ``hidden symmetries" of 11-dimensional supergravity or an appropriate extension \cite{Julia:1980gr,Julia:1982gx,Julia1,Nicolai:1991kx,Julia:1997cy,West:2001as,Damour:2000hv,Damour:2002cu}.  Weyl reflections in the exceptional root may change the spacetime signature.   In the orbit of the standard signature $(10,1, +)$ of $M$-theory (the one on which we shall focus in this paper) appear also the signatures $(9,2,-)$ (``$M^*$-theory"), $(6,5,+)$ (``$M'$-theory"), and the theories with reversed signatures $(1,10, -)$, $(2,9, +)$ and $(5,6, -)$ \cite{Englert:2003py,Keurentjes:2004bv,Keurentjes:2004xx,Englert:2004ph}.  In $(s,t, \pm)$, $s$ is the number of space directions, $t$ the number of time directions, and  $\pm$ refers to the sign of the kinetic term of the $3$ form. The analysis can be extended to other hidden symmetry algebras \cite{deBuyl:2005it}.  Multiple time physics has a long history (for a survey of two-time physics, see \cite{Bars:2000qm}) and has recently been shown to be connected with supergroup gauge theories and negative branes \cite{Dijkgraaf:2016lym}.  

For all these reasons, it is of interest to explore solutions with exotic spacetime signatures in the context of $M$-theory. This is the object of this note. The analogues of the Freund Rubin solutions \cite{Freund:1980xh} have been already studied in \cite{Hull:1998fh}. We derive here the analogue of the Englert solution \cite{Englert:1982vs}.

It is known that the seven-sphere $S^7 \equiv SO(8)/SO(7)$ is one of the only spheres that are parallelizable \cite{Cartan1,Cartan2,Adams}.   If one allows pseudo-riemannian signatures, the family includes also the pseudo-sphere $S^{3,4}$  \cite{Wolf1, Wolf2},  which is the space of constant curvature  $SO(4,4)/SO(3,4)$ or space of unit split octonions  (see also \cite{Ranjbar})\footnote{Of course, the negative curvature spaces obtained by multiplying the metric by an overall minus sign, which are the pseudo-hyperbolic spaces $H^{0,7}$ and $H^{4,3}$  (see appendix \ref{app-notation} for conventions), also  enjoy this property.}.   Now, the seven-sphere plays an important role in eleven-dimensional supergravity \cite{Cremmer:1978km}, either through the Freund-Rubin solution where it leads to the $SO(8)$ gauged supergravity  \cite{deWit:1982bul} or through the Englert solution \cite{Englert:1982vs} which breaks the $SO(8)$ symmetry down to $Spin(7)$ \cite{Englert:1983jc} through the turning on of internal fluxes.   We show that there is an analogue of the Englert solution in $M'$-theory, which takes the form $AdS_4 \times S^{3,4}$.  Just as its original parent, this exotic form of the Englert solution is characterized by internal fluxes provided by a torsion that parallelizes $S^{3,4}$.  These internal fluxes break the $SO(4,4)$ symmetry of the pseudo-sphere $S^{3,4} \equiv SO(4,4)/SO(3,4)$ down to $Spin^+(3,4)$ (see Appendix \ref{App:pseudo} for information on the pseudo-sphere $S^{3,4}$, the split octonions and the relevant associated groups).

\section{Action principle and general ansatz}

\subsection{Action}

Our starting point is the bosonic action of $11$-dimensional supergravity with mixed signature \cite{Cremmer:1978km,Hull:1998ym, Hull:1998vg}, written as
\be
S=\f{1}{2\kappa^2} \int dx^{s+t} \mathcal{L},
\ee
where the Lagrangian $\mathcal{L}$ is
\be
\mathcal{L}=\sqrt{|g^{s+t}|} (R-\f{\eta}{24}F_{MNPQ}F^{MNPQ}-\f{2\sqrt{2}}{12^4}\f{1}{\sqrt{|g^{s+t}|}}\varepsilon^{M_1...M_{11}}F_{M_1...M_4}F_{M_5...M_8}A_{M_9...M_{11}}).
\ee
Here, $s+t$-dimensions ($s+t=11$) refers to $s$ spacelike directions and $t$ timelike directions and $F_{MNPQ}=4\partial_{[M}A_{NPQ]}$. The sign of $\eta$ is determined as\footnote{From the point of view of fermionic content of the theory, whenever $\eta=+1$ there exists a purely real representation of Clifford algebra and as a result spinors are Majorana. However, if $\eta=-1$ then there exists a purely imaginary representation of Clifford algebra and the spinors are pseudo-Majorana \cite{Kugo:1982bn}.}
\begin{align}
\eta &= +1\quad \textrm{if}\quad (s-t)\, \textrm{mod}\,8=1\quad\textrm{i.e. in dimensions}\quad 10+1, 6+5, 2+9,\\
\eta &= -1\quad \textrm{if}\quad (s-t)\, \textrm{mod}\,8=7\quad\textrm{i.e. in dimensions}\quad 1+10, 5+6, 9+2.
\end{align}
The equations of motion are
\begin{align}
R_{MN}-\f{1}{2}g_{MN}R &= -\f{\eta}{48}(g_{MN}F^2-8 F_{MPQR}F\indices{_N^{PQR}}),\label{EOM:var-wt-metric}\\
F\indices{^{MNPQ}_{;M}} &= \f{1}{\eta} \f{18\sqrt{2}}{12^4}\f{1}{\sqrt{|g^{11}|}} \varepsilon^{M_1...M_8 NPQ}F_{M_1...M_4}F_{M_5...M_8}\label{EOM:var-wt-3form}.
\end{align}

As we mentioned in the introduction, $M$-theory corresponds to $\eta = 1$, $(s,t)= (10,1)$, whereas $M^*$-theory corresponds to $\eta = -1$, $(s,t) = (9,2)$ and $M'$-theory corresponds to $\eta = 1$, $(s,t) = (6,5)$.  The equations of motion are invariant under the sign change of the metric $g_{MN} \rightarrow - g_{MN}$ accompanied by $\eta \rightarrow - \eta$ as well as $A_{MNP} \rightarrow  A_{MNP}$.  We recall that with our conventions,  $\epsilon^{01\cdots 10}$ changes sign since in the transformation $g_{MN} \rightarrow - g_{MN}$,  the parity of the number of time directions changes.  Because of this invariance, we concentrate on $M$-theory, $M^*$-theory and $M'$-theory,  leaving to the reader  the trivial task of deriving the solutions for the reversed signature theories corresponding to the other theories in the same duality orbit as ordinary supergravity.

\subsection{Product of constant curvature spaces}

We assume that the eleven-dimensional spacetime manifold (referred to as the ``ambient space" in the sequel) splits as the product of a four-dimensional manifold (``background spacetime") and a seven-dimensional manifold (``internal space")\footnote{We stress that the ``internal space" is thus in our terminology always the seven-dimensional manifold, independently of its curvature or signature.}.  The metric itself is a direct sum and splits into two parts, the metric on the internal space, and the one on the background spacetime. Capital Latin indices $M,N,...$ run over $0,...,10$. Small Latin indices $m,n,...$ are internal space indices and run over $4,...,10$.  Greek indices $\mu,\nu,...$ are background spacetime indices running over $0,1,2,3$. Notice that here $|g^{11}|$ is the absolute value of the determinant of the metric of the total space. We use the absolute value everywhere when needed for the determinant of the ambient, internal and background metrics since, a priori, there is no restriction on the signature of the aforementioned metrics. 

We also assume that spacetime and the internal space are spaces of constant curvature, i.e., either a pseudo-sphere or a pseudo-hyperbolic space, with curvatures respectively given by 
\begin{align}
 R_{mnrs} &= \f{a}{6} \left(g_{mr} g_{ns} - g_{ms} g_{nr} \right) \label{CC1},\\ 
R_{\mu \nu \rho \sigma} &= \f{b}{3} \left(g_{\mu \rho} g_{\nu \sigma} - g_{\mu \sigma} g_{\nu \rho} \right) \label{CC2},
\end{align}
where $a$ and $b$ are constants, which are positive for pseudo-spheres and negative for pseudo-hyperbolic spaces.  The conventions and notations used in this paper are collected in Appendix \ref{app-notation}.  This implies in particular
\begin{align}
R_{mn} &=a\, g_{mn},\label{Ricci-internal-space}\\
R_{\mu\nu} &=b \, g_{\mu\nu},\label{Ricci-spacetime}\\
R_{m\mu} &=0.\label{Ricci-mixed}
\end{align}

\section{Freund-Rubin type solutions}

We first derive the analogues of the Freund-Rubin solutions.  These were already considered in \cite{Hull:1998fh}.

The Freund-Rubin ansatz assumes in addition to the above conditions on the metric that the only $4$-form flux is in spacetime and reads explicitly: 
\begin{align}
F^{\mu\nu\rho\sigma} &=\f{1}{\sqrt{|g^4|}}\f{f}{\sqrt{4!}} \varepsilon^{\mu\nu\rho\sigma},\label{flux-background}\\
F^{mPQR} &=0,
\end{align}
where $f$ is a constant. There is no internal flux.  

Now, from \eqref{EOM:var-wt-metric} and  the ansatz,  one finds that
\begin{align}
R_{mn} &= (-\f{1}{3}) \f{\eta}{24}f^2 (-1)^{T'}  g_{mn},\\
R_{\mu\nu} &= (+\f{2}{3}) \f{\eta}{24}f^2 (-1)^{T'}  g_{\mu\nu},
\end{align}
where $T'$ is the number of timelike directions in the four-dimensional background spacetime metric.  We shall also introduce $T$ as the number of timelike directions in the internal space, so that $t$  (the number of timelike directions of the eleven-dimensional ambient space) is equal to $t = T + T'$. Comparing with \eqref{Ricci-internal-space} and \eqref{Ricci-spacetime}, we find
\be\label{FR-solution}
a=-\f{1}{3}\f{\eta}{24}f^2 (-1)^{T'}, \qquad b=-2a,
\ee
as in \cite{Hull:1998fh}.
For example, the standard Freund-Rubin solution is obtained by setting $\eta=+1$, $T'=1$.

The Freund-Rubin solutions constitute a one-parameter family of solutions.  One can take $f$ as the free parameter characterizing the solutions.  The curvature of the background spacetime and of the internal space are then determined by $f$ and are of opposite signs.  Note that they depend in fact on $f^2$ and so are invariant under $f \rightarrow -f$ as it should. 

The sign of $a$ is easily seen to be always equal to  $(-1)^T$: for  $M$-theory and $M'$-theory, one has $\eta = 1$ and $(-1)^{T'} =-(-1)^T$ because $T+T' = t $ is odd.  This is also true for the reversed $M^*$-theory with signature $(2,9)$ and $\eta = 1$. For  $M^*$-theory, one has $\eta = -1$ but $t$ is now even and then $(-1)^{T'} =(-1)^T$.  The same is true for the reversed signature $M$- and $M'$-theories.  In all cases, the curvatures in the four-dimensional spacetime and in the internal manifold have opposite signs.

All Freund-Rubin solutions that can be obtained in this way are maximally supersymmetric \cite{Hull:1998fh}.

\section{Englert type solutions}

\subsection{Ansatz}

The Englert type solutions \cite{Englert:1982vs} can be considered as spontaneous breakings of the Freund-Rubin solutions  through non-vanishing expectation values for some of the  internal components of the $4$-form.  In other words there is now an internal flux which breaks completely the supersymmetry. 

The Englert construction is available when the internal manifold is parallelizable.  This occurs when the internal  space of constant curvature is either the sphere $S^{7,0}$, corresponding to the original Englert solution \cite{Englert:1982vs}, or the pseudo-sphere $S^{3,4}$ (as well, of course, as the cases trivially obtained from these ones by an overall change of sign of the metric, corresponding to the pseudo-hyperbolic spaces $H^{0,7}$ and $H^{4,3}$).  All the other cases ($S^{6,1}$, $S^{5,2}$,  $H^{6,1}$ etc) are not parallelizable.  

By matching the signatures, one easily sees that there are a priori four cases where the parallelizable seven-(pseudo-)spheres might appear for the theories in the time-duality orbit of $M$-theory.  These are
\begin{itemize}
\item Case 1: $M$-theory with spacetime manifold  $M^{3,1} \times S^{7,0}$,
\item Case 2: $M^*$-theory with spacetime manifold $M^{2,2} \times S^{7,0}$,
\item Case 3: $M'$-theory with spacetime manifold $M^{3,1} \times S^{3,4}$,
\item Case 4: reversed $M'$-theory with spacetime manifold $M^{2,2} \times S^{3,4}$,
\end{itemize}
as well as the reversed signature solutions.  Here, $M^{p,q}$ is a priori either $S^{p,q}$ or $H^{p,q}$, i.e., we leave the sign of the curvature of four-dimensional spacetime open. It will be determined by the equations.  We shall show that only cases 1 and 3 are actually compatible with the equations of motion and furthermore, that $M^{3,1}$ is then the anti-de Sitter space $H^{3,1}$.

We note that in all cases, the product $\eta (-1)^{T'}$, where $T'$ is the number of time directions in the four-dimensional background spacetime, is equal to $-1$,
$$\eta (-1)^{T'} = -1.$$
Indeed, one has $\eta = 1$ for cases 1 and 3 with $T'$ odd, and $\eta = -1$ for cases 2 and 4 but $T'$ is now even.  Furthermore, the curvature $a$ of the internal space is positive,
$$a >0,$$
and the number of time directions in the internal space is even ($a$ would of course be negative and $T$ odd for the reversed signature solutions.)

The Englert ansatz \cite{Englert:1982vs} consists in imposing the earlier conditions (\ref{CC1}), (\ref{CC2}),  (\ref{flux-background}) and
\be
F^{m\mu QR} =0. \label{flux-mixed}
\ee
The condition that the internal flux should vanish is, however, dropped, i.e., one allows $F^{mnpq}  \neq 0$. 
In fact, one postulates instead 
\be
F_{mnpq}=\lambda S_{[npq,m]} = \lambda S_{npq;m}. \label{flux-internal}
\ee
where $S_{npq}$ is one of the (metric compatible) torsions that parallelizes the internal space, i.e., such that when added to the Levi-Civita torsion-free connection,   the resulting connection is 
flat (zero curvature).  In (\ref{flux-internal}),   $\lambda$ is a constant parameter.

As Cartan and Schouten showed \cite{Cartan1, Cartan2} (see also \cite{Englert:1982vs}), the torsions satisfy the following identities:
\begin{align}
S\indices{^{tr}_m} S_{trn} &=a\, g_{mn}, \label{Torsion101}\\
S\indices{_{tm}^r} S\indices{_{rn}^s} S\indices{_{sp}^t} &=\f{1}{2}a\, S_{mnp}, \label{Torsion102}\\
S_{npq;m} = S_{t[np}S\indices{^t_{q]m}} &= S_{[npq,m]}.  \label{Torsion103}
\end{align}
The parallelisms are respectively related to the octonion and split octonion algebras.  Because these algebras are non-associative, there are in each case two infinite classes, denoted $+$ and $-$.  We refer to \cite{Rooman} for useful information concerning  the $S^{7,0}$ sphere, easily extendable to $S^{3,4}$.   The corresponding torsions can be expressed in terms of the structure constants of the octonion or split-octonion algebras.   The invariance group at any given point of these $3$-forms is isomorphic to $G_2$ ($G^*_{2,2}$) in the Riemannian 
(pseudo-Riemannian) case (see Appendix \ref{App:pseudo}). The fact that there are only two possible internal manifolds $S^{7,0}$ or $S^{3,4}$, and that $S^{6,1}$, $S^{5,2}$ etc are excluded,  is connected with the fact that there are only two real forms of the octonion algebra, and two real forms of the Lie algebra $g_2$.

Furthermore, one finds from (\ref{Torsion101})-(\ref{Torsion103}) that the
torsion $3$-forms are eigenfunctions of the (pseudo-)Laplace operator in $7$ dimensions,  
\be
S\indices{_{npq;m}^{;m}}=-\f{2}{3} a \, S_{npq}. \label{Laplace}
\ee
One can also prove that the torsions are dual or anti-dual to their curvature \cite{Englert:1982vs},
\begin{align}
S^{mnp} &= \pm \sqrt{\frac{6}{a}}\frac{1}{4!}\frac{1}{\sqrt{|g^7|}}\varepsilon^{mnpqrst}S_{[rst,q]},\\
S_{mnp} &= \pm \sqrt{\frac{6}{a}}\frac{1}{4!}\sqrt{|g^7|}\varepsilon_{mnpqrst}S^{[rst,q]}.
\end{align}
where the $\pm$ sign depends on the $\pm$-class to which the torsion belongs.

With the above assumptions, the equation \eqref{EOM:var-wt-3form} for $F_{MNPQ}$ reduces to
\be\label{EOM-for-F-Englert}
F\indices{_{mnpq}^{;m}}= \f{(-1)^{T'}}{\eta} \f{f}{2(12)^{3/2}}\sqrt{|g^{7}|}\,\varepsilon_{rstunpq}F^{rstu},
\ee
 and is therefore solved in view of (\ref{flux-internal}) and (\ref{Laplace}) if we impose
\be\label{alpha-for-Englert-sol}
a = -\f{3}{4}\f{\eta}{24}f^2 (-1)^{T'} = \f{1}{32}f^2.
\ee
Given $a$, the two choices $f=\pm |f|= \pm \sqrt{32 a}$  correspond to the $+$ or $-$ parallelism, respectively.

Now, the equation \eqref{EOM:var-wt-metric} is equivalent to 
\be
R_{MN} = -\f{\eta}{48}\left(\f23 g_{MN}F^2-8 F_{MPQR}F\indices{_N^{PQR}}\right),\label{EOM:var-wt-metric2}
\ee
Using the form of the fields obtained so far, the $(mn)$-components of \eqref{EOM:var-wt-metric2} read
\be
a = \frac49 a + \frac{1}{24}\frac{10}{9} \eta \lambda^2 a^2,
\ee
yielding
\be
\lambda^2 = \eta \frac{12 }{a}.
\ee
Since $a>0$, this equation has no (real) solution when $\eta = -1$, i.e., for cases 2 and 4 above.   When $\eta = 1$, this equation determines $\lambda$ in terms of $a$,
\be
\lambda = \pm \sqrt{\frac{12}{a}}.
\ee
The $(\mu \nu)$-components of \eqref{EOM:var-wt-metric2} determine then the radius of curvature $b$ of the four-dimensional background spacetime,
\be
b = - \frac53 a,
\ee
which is thus negative.  The four-dimensional background spacetime is therefore anti-de Sitter space $AdS_4 \equiv H^{3,1}$.

\subsection{Summary of solutions}

We thus see that there are only two cases among the cases listed above that can actually be realized (and the cases obtained by reversing the signature). These are
\begin{enumerate}
\item The Englert solution $ H^{3,1} \times S^{7,0}$ in $M$-theory.
\item The ``exotic" Englert solution  $H^{3,1} \times S^{3,4}$  in $M'$-theory.
\end{enumerate}
For all these solutions $\lambda^2=\f{12}{a}$ and $b=-\f{5}{3} a$, in consistency with \cite{Englert:1982vs}.

The parallelizable seven-sphere or seven-pseudo-sphere with internal torsion (fluxes) turned on is therefore a possible solution only in $M$-theory and $M'$-theory. $M^*$-theory does not accomodate such solutions.  Furthermore, the four-dimensional background spacetime is in both cases anti-de Sitter, with a single time.  There is no solution with two time directions in the four-dimensional background spacetime. 

\subsection{Breaking of symmetry}
The Englert solution of $M$-theory breaks both supersymmetry and the $SO(8)$-symmetry of the seven-sphere down to $Spin(7)$ \cite{Englert:1983jc}.

A similar situation prevails in $M'$-theory.
The internal fluxes of the Englert-type solution break the $SO(4,4)$ symmetry of the pseudo-sphere $S^{3,4}$.  This is because the $4$-form $F_{mnpq}$ at any given point is not invariant under the full isotropy subgroup $SO(3,4)$.  It is only invariant under the subroup $G^*_{2,2}$ of the split octonions, since the torsion is determined by the structure constants of the split octonions.  Now, the pseudo-sphere is also equal to $Spin^+(3,4)/G^*_{2,2}$, where the group $Spin^+(3,4)$ is the connected component of $Spin(3,4)$ (see \cite{Kath:1996gm} and Appendix \ref{App:pseudo}).  This implies that the symmetry group of the $M'$-theory Englert solution is $Spin(3,4)$.  Similarly,  supersymmetry is broken.


\section{Conclusions and comments}

If one adopts the idea suggested by duality that the spacetime signature is a relative concept that depends on the description \cite{Hull:1998ym, Hull:1998vg}, it is natural to explore solutions of the exotic forms of $M$-theory in which there is more than one time direction.  We have shown in this note that an exotic version of the Englert solution exists in $M'$-theory, where the internal manifold is the parallelizable pseudo-sphere $S^{3,4}$. Through the non-vanishing internal fluxes, this solution breaks the $SO(4,4)$ symmetry of the pseudo-sphere down to $Spin^+(3,4)$. 

It is interesting to note that $M$-theory and $M'$-theory have differences $s-t$ of the number of space directions minus the number of time directions that are equal modulo $8$ ($10-1 = 6-5$ modulo $8$).  Accordingly, they have spinor representations with similar properties. In fact, the spinor representations of $SO(7)$ and $SO(3,4)$  are in both cases $8$-dimensional and real.

The pseudo-sphere $S^{3,4}$ admits Killing spinors in terms of which one can express the torsions \cite{Kath:1996gm}.  Deformation by squashing of the pseudo-sphere can aso be contemplated.

\section*{Acknowledgments} 
This work was partially supported by the ERC Advanced Grant ``High-Spin-Grav" and by FNRS-Belgium (convention FRFC PDR T.1025.14 and  convention IISN 4.4503.15).
 \break

\begin{appendix}
\section{Notation}\label{app-notation}
In this paper, we generally follow the notation of \cite{Freedman}.  For concreteness, we spell out the needed ones here:
\begin{itemize}
\item
$s$: Number of space directions of the ambient space manifold

\item
$t$: Number of time directions of the ambient space manifold

\item
$T$: Number of time directions of the internal manifold

\item
$T'$: Number of time directions of spacetime ``background" manifold

\item
$g_{MN} =\left( \begin{smallmatrix} g_{\mu\nu} & 0 \\  0 & g_{mn} \end{smallmatrix} \right)$: the metric on $11$ dimensional ambient space, $M,N=0,...,10$.

\item
$g_{mn}$: the metric on $7$ dimensional internal space, $m,n=4,...,10$.

\item
$g_{\mu\nu}$: the metric on $4$ dimensional spacetime, $\mu,\nu=0,...,3$.

\item
$F_{MNPQ}=4 \partial_{[M}A_{NPQ]}$.

\item
$\varepsilon_{M_1...M_{11}}$: the $11$ dimensional Levi-Civita antisymmetric covariant tensor density of rank $11$ and weight $-1$,  with $\varepsilon_{01234...}=+1$.

\item
$\varepsilon^{M_1...M_{11}}$: the $11$ dimensional Levi-Civita antisymmetric contravariant tensor density of rank $11$ and weight $+1$,  with $\varepsilon^{01234...}=(-1)^t$.  This choice is such that
$$\f{1}{\sqrt{\vert g^{11}\vert}}\varepsilon^{M_1 ... M_p}= \sqrt{\vert g^{11}\vert}  g^{M_1 N_1} g^{M_2 N_2}\cdots g^{M_{11} N_{11}} \varepsilon_{N_1 ... N_p},$$  where $g^{11}$ is the determinant of $g_{MN}$.  

\item
\begin{align}
\varepsilon_{M_1...M_k M_{k+1}...M_{k+p}}\varepsilon^{M_1...M_k N_{k+1}...N_{k+p}}&=(-1)^t\, k!\,\delta^{N_{k+1}...N_{k+p}}_{M_{k+1}...M_{k+p}}\nonumber\\
&=(-1)^t\, k!\, p!\, \delta^{N_{k+1}}_{[M_{k+1}}\delta^{N_{k+2}}_{M_{k+2}}...\delta^{N_{k+p}}_{M_{k+p}]}.\nonumber
\end{align}

\item
$SO(p,q)$: the group of all transformations which leaves invariant the bilinear form $\eta_{p,q}=\sum_{i=1}^{p} dx_i^2 - \sum_{j=1}^{q} dx_j^2$.

\item
The pseudo-sphere $S^{p-1,q}=SO(p,q)/SO(p-1,q)$ ($p \geq 1, q \geq 0$) is a manifold with induced metric of  signature $(p-1,q)$ and  positive curvature.  In particular, $S^{p-1,0}$ is the standard $(p-1)$-sphere $S^{p-1}$.

\item
The  pseudo-hyperbolic space $H^{p,q-1}=SO(p,q)/SO(p,q-1)$ ($p \geq 0, q \geq 1$) is a manifold with induced metric of  signature $(p,q-1)$ and negative curvature.  While one removes a spacelike direction from the ambient space to get $S^{p-1,q}$, one removes a timelike direction for $H^{p, q-1}$ (see Appendix \ref{App:pseudo} for an explicit example).  In particular, $H^{p,0}$ is the standard hyperbolic space $H^p$ of dimension $p$. 

\end{itemize}

\section{The pseudo-sphere $S^{3,4}$ and the split octonions}
\label{App:pseudo}

\subsection{The pseudo-sphere $S^{3,4}$ as a homogeneous space}
Let $\mathbb{R}^{4,4}$ be the $8$-dimensional real vector space endowed with the flat metric of mixed signature $(4,4)$,
\be
ds^{2}_{4,4} = \left(dx^1\right)^2 + \left(dx^2\right)^2 + \left(dx^3\right)^2 + \left(dx^4\right)^2 - \left(dy^1\right)^2 -\left(dy^2\right)^2 - \left(dy^3\right)^2 - \left(dy^4\right)^2. 
\ee
We consider the  hypersurfaces:
\begin{eqnarray}
S^{3,4} &:& \left(x^1\right)^2 + \left(x^2\right)^2 + \left(x^3\right)^2 + \left(x^4\right)^2 - \left(y^1\right)^2 -\left(y^2\right)^2 - \left(y^3\right)^2 - \left(y^4\right)^2 = 1,\\
H^{4,3} &:& \left(x^1\right)^2 + \left(x^2\right)^2 + \left(x^3\right)^2 + \left(x^4\right)^2 - \left(y^1\right)^2 -\left(y^2\right)^2 - \left(y^3\right)^2 - \left(y^4\right)^2 = -1.
\end{eqnarray}
These are connected. Because the signature of the embedding space is symmetric for the exchange of the space and time directions, these hypersurfaces are clearly isomorphic.  The ``pseudo-sphere" $S^{3,4}$ has an induced metric with signature $(3,4)$ (three $+$ signs and four $-$ signs), while the ``pseudo-hyperbolic space" $H^{4,3}$ has an induced metric with signature $(4,3)$.   One goes from one to the other by an overall sign change of the metric.  

The split group $O(4,4)$ and its subgroups $SO(4,4)$ and $SO^+(4,4)$ act transitively on $S^{3,4}$ and $H^{4,3}$.  The stability subgroups at a point are respectively $O(3,4)$, $SO(3,4)$ and $SO^+(3,4)$ for $S^{3,4}$, and $O(4,3)$, $SO(4,3)$ and $SO^+(4,3)$ for $H^{4,3}$.  Of course, the groups $O(3,4)$ and $O(4,3)$ (as well as $SO(3,4)$ and $SO(4,3)$, or  $SO^+(3,4)$ and $SO^+(4,3)$) are isomorphic, but we prefer to adopt different notations in each case to keep track of the signature of the metric.  Thus, the pseudosphere $S^{3,4}$ and pseudo-hyperbolic space $H^{4,3}$ are the homogeneous spaces:
\be 
S^{3,4} = \frac{O(4,4)}{O(3,4)} = \frac{SO(4,4)}{SO(3,4)} =  \frac{SO^+(4,4)}{SO^+(3,4)},
\ee
and
\be
H^{4,3} = \frac{O(4,4)}{O(4,3)} = \frac{SO(4,4)}{SO(4,3)} =  \frac{SO^+(4,4)}{SO^+(4,3)}.
\ee

The pseudo-sphere $S^{3,4}$ and pseudo-hyperbolic space $H^{4,3}$ are not only homogeneous spaces, they are in fact maximally symmetric and hence spaces of constant curvature,
\be R_{mnpq} = K \left(g_{mp}g_{nq} - g_{mq}g_{np} \right),
\ee
with $K=1 >0$ for the pseudo-sphere $S^{3,4}$ and $K = -1 <0$ for the pseudo-hyperbolic space $H^{4,3}$,  in agreement with the observation that under an overall change of sign of the metric, the Riemann curvature changes sign but the product $\left(g_{mp}g_{nq} - g_{mq}g_{np} \right)$ does not, so that $K$ changes sign.

\subsection{Split octonions}

Let $a$ and $b$ be two quaternions, $a = a_0 + a_1 i + a_2 j + a_3 k$, $b = b_0 + b_1 i + b_2 j + b_3 k$ ($a_i, b_i \in \mathbb{R}$, $i^2 = j^2 = k^2 = -1$, $ij = k = - ji$ etc). The split octonions are pairs of quaternions $a +   b \ell $, where $\ell $ is a new element, for which one defines the product as
\be 
(a +  b \ell) (c +  d \ell ) = ac + (\ell)^2 \bar{d} b +  (d a + b \bar{c}) \ell, \label{Product}
\ee
where
\be \ell^2 = +1,
\ee
instead of $-1$ as for the standard octonions.  In (\ref{Product}), the overbar denotes quaternionic conjugation, $\bar{d} = d_0 - d_1 i - d_2 j - d_3 k$.

Setting $e_0 = 1$, $e_1 = i$, $e_2 = j$, $e_3 = k$, $e_4 = \ell$, $e_5 = i \ell$, $e_6 = j \ell$ and $e_7 = k \ell$, one gets the same products $e_i e_j$ as for the standard octonions, except when two $\ell$'s are involved, in which case one gets an additional minus sign. For instance $e_1 e_4 = e_5$ as for octonions, but $e_4 e_5 = -e_1$ while it is $+e_1$ for standard octonions.  Similarly, $e_4^2 = e_5^2 = e_6^2 = e_7^2 = +1$ instead of $-1$.

One defines octonionic conjugation for a general split octonion
\be
x =  x^0 e_0 + x^1 e_1 + x^2 e_2 + x^3 e_3 + x^4 e_4 + x^5 e_5 + x^6 e_6 + x^7 e_7 , \; \; \; x^i \in \mathbb{R}
\ee
as
\be
\bar{x}  =  x^0 e_0 - x^1 e_1 - x^2 e_2 - x^3 e_3 - x^4 e_4 - x^5 e_5 - x^6 e_6 - x^7 e_7.
\ee
One has
\be
\overline{xy} = \bar{y} \bar{x},
\ee
and of course $\overline{\bar{x}} = x$.
A scalar product can then be introduced as
\be
(x,y) = \frac12 \left( x \bar{y}+ y \bar{x} \right) = \frac12 \left( \bar{x} y+ \bar{y} x \right).
\ee
Explicitly, one finds
\begin{eqnarray}
(x,y) &=& x^0 y^0 + x^1 y^1 + x^2 y^2 + x^3 y^3 - x^4 y^4 - x^5 y^5 - x^6 y^6 - x^7 y^7 \\
&=& x^0 y^0 + \sum_{i=1}^7 \eta_{ij} x^i y^j,
\end{eqnarray}
where $\eta_{ij}$ is the flat metric with signature $(3,4)$.

An octonion is pure imaginary if $\bar{x} = - x$.  The unit octonions $e_i$ are pure imaginary.  The pure imaginary condition is equivalent to  $(e_0, x) = 0$. 

One can rewrite the product of the unit octonions $e_i$ as
\be
e_i e_j = - \eta_{ij} e_0+ a_{ij}^{\; \; \; k} e_k,
\ee
where the structure constants $a_{ij}^{\; \; \; k}$ are such that the $a_{ijk} \equiv a_{ij}^{\; \; \; m} \eta_{km}$ are completely antisymmetric in $(i,j,k)$.   The tensor $a_{ijk}$ has value $+1$ for $(ijk) = (123)$, $(154)$, $(167)$, $(264)$, $(275)$, $(374)$ and $(356)$ (and cyclic permutations).

The squared norm $N(x) \equiv (x,x)$ of $x$ reads
\begin{eqnarray}
N(x)  &=& \left( x^0 \right)^2 +  \left( x^1 \right)^2 +  \left( x^2 \right)^2 +  \left( x^3 \right)^2 -  \left( x^4 \right)^2 - \left( x^5 \right)^2 -  \left( x^6 \right)^2 -  \left( x^7 \right)^2 \\
&=& \eta_{\lambda \mu} x^\lambda x^\mu, \; \; \; \; (\lambda, \mu = 0, 1, \cdots, 7)
\end{eqnarray}
where $\eta_{\lambda \mu}$ is the flat metric with signature $(4,4)$.  The split octonions form a composition algebra, i.e., 
\be
N(xy) = N(x) N(y).
\ee

Just as the standard octonions, the split octonions do not form an associative algebra.  However, in the same way as for the standard octonions, the associator $[x,y,z]$ of three split octonions $x$, $y$, $z$, defined through
\be
[x,y,z] = (xy)z - x(yz),
\ee 
is an alternating function of $x$, $y$, $z$, i.e. 
\be
[x,y,z] = [x,y,z] = [z,x,y] = - [y,x,z] = -[x,z,y] = -[z,y,x].
\ee

More information on the split octonions can be found in \cite{Gunaydin:1973rs,BaezHuerta}.

\subsection{Parallelizations of the pseudo-sphere $S^{3,4}$}

There exist two infinite families of parallelizations of the pseudo-sphere $S^{3,4}$.  We first start by describing one of these parallelizations.

Consider the pseudo-sphere $S^{3,4}$ of unit octonions, $x \in S^{3,4} \Leftrightarrow N(x) =1$.  The real number $1$ is the ``North pole".  The tangent space at the North pole can be identified with the seven-dimensional vector space of imaginary octonions.  A Lorentz basis of this tangent space is given by the $e_i$'s.  Let  $x \in S^{3,4}$.   Right multiplication by $x$ maps  the North pole to $x$, and the tangent space at the North pole to the tangent space at $x$.  Indeed, the vectors $e_i$ at the North pole are mapped on $t^+_i  \equiv e_i x$.  One has:
\be
(t^+_i, x) = 0, \; \; \; (t^+_i, t^+_j) = \eta_{ij}.
\ee
The first equality expresses that the seven octonions $t^+_i$, viewed as vector in $\mathbb{R}^{4,4}$ are orthogonal to $x$ and hence tangent to the pseudo-sphere $S^{3,4}$ at $x$.  The second equality expresses that the $t^+_i$'s form a Lorentz basis of that tangent space.  

We have thus defined at each point of the pseudo-sphere $S^{3,4}$ a Lorentz basis of the tangent space.  This provides an absolute parallelism for $S^{3,4}$. The tangent vector at $x$ parallel to the tangent vector $e_i$ at $1$ is the vector $e_i x$ obtained by right multiplication with $x$.  The corresponding parallel transport preserves the metric of $S^{3,4}$ and its geodesics can be verified to coincide with those defined by the metric. The parallelization defined by $\{t^+_i\}$ is thus consistent with the metric. 

Similarly, left multiplication also defines a parallelization of $S^{3,4}$ that maps the tangent basis $\{e_i\}$ at $1$ on the tangent basis $\{t_i^- \equiv x e_i\}$ at $x$.  The two parallelisms are inequivalent since the tangent vectors $t^+_i$ and $t_i^-$   coincide only at the North pole and at the ``South pole" $-1$. 

Yet other parallelizations can be defined by using a reference point on  $S^{3,4}$ different from unity.  More precisely, let $\alpha \in S^{3,4}$.  One goes from $\alpha$ to $x$ by multiplying $\alpha$ by $\bar{\alpha} x$, $\alpha (\bar{\alpha } x) = x$.  One defines the tangent vector $^{(\alpha)} t_i^+$ at $x$ parallel to the tangent vector $e_i \alpha$ at $\alpha$ through right multiplication by the octonion $\bar{\alpha} x$ that connects $\alpha$ to $x$, $^{(\alpha)} t_i^+ = (e_i \alpha) (\bar{\alpha} x)$.  The tangent vectors $^{(\alpha)} t_i^+$ and $t^+_i$ at $x$ do not coincide because octonionic multiplication is not associative.  A similiar construction yields the parallelism $^{(\alpha)} t_i^- = (x \bar{\alpha})(\alpha e_i)$.

Because the families of parallelisms given by the above construction are consistent with the metric, the corresponding torsion tensors obey the equations (\ref{Torsion101})-(\ref{Torsion103}) given above \cite{Cartan1,Cartan2,Wolf1,Wolf2}.  

The parallelisms of $S^{3,4}$ are related to the split octonions in the same way as the parallelisms of the seven-sphere are related to the standard octonions.  For that reason, the reader can find more information on the parallelisms of $S^{3,4}$ in the literature on the parallelisms of the seven-sphere.  A reference that we have found useful is \cite{Rooman}.

\section{$Spin^+(3,4)$ and $G_{2,2}^*$}

The complex Lie agebra $g_2$ possesses two real forms, the compact one and the split one.  To the compact real form corresponds the unique compact group $G_2$.  To the split real form correspond the simply connected non compact group $G_{2,2}$ with center $\mathbb{Z}_2$ and the quotient $G_{2,2}^* \equiv \frac{G_{2,2}}{\mathbb{Z}_2} $ (``adjoint real form") which has trivial center (and is not simply connected). The group  $G_{2,2}^*$ is the automorphism group of the split octonions.

The group $Spin(7)$ is well-known to have a transitive action on the seven-sphere $S^7$, with isotropy group $G_2$.  Similarly, the group $Spin^+(3,4)$ (connected component of $Spin(3,4)$) has a transitive action on the pseudo-sphere $S^{3,4}$ with isotropy group $G_{2,2}^*$ \cite{Kath:1996gm}.  We can thus also identify $S^{3,4}$ with the homogeneous space $Spin^+(3,4) /G_{2,2}^*$,
\be
S^{3,4} \simeq \frac{Spin^+(3,4)}{G_{2,2}^*}
\ee

\end{appendix}


\end{document}